\documentclass[11pt]{article}
\usepackage{epsfig}

\newcommand{\beq}{\begin{equation}}
\newcommand{\eeq}{\end{equation}}
\newcommand{\beqa}{\begin{eqnarray}}
\newcommand{\eeqa}{\end{eqnarray}}

\newcommand{\tr}{{\rm Tr}}

\begin{document}

\begin{titlepage}
\def\thepage {}        % Kill page numbering

\title{Two-gluon coupling and collider phenomenology of color-octet technirho 
mesons}

\author{
R. Sekhar Chivukula$^1$, Aaron Grant$^2$, Elizabeth H. 
Simmons$^{1,2,3}$\thanks{e-mail addresses: sekhar@bu.edu, 
grant@carnot.harvard.edu, simmons@bu.edu}\\
  \\
$^1$Department of Physics, Boston University, \\
590 Commonwealth Ave., Boston MA  02215 \\
$^2$Jefferson Laboratory of Physics, Harvard University\\
Cambridge MA,  02138 \\
$^3$Radcliffe Institute for Advanced Study, Harvard University\\
34 Concord Avenue, Cambridge MA  02138}

\date{}

\maketitle

\bigskip
\begin{picture}(0,0)(0,0)
\put(295,250){BUHEP-01-19}
\put(295,235){HUPT-01/A042}
\put(295,220){hep-ph/0109029}
\end{picture}
%\vspace{-12pt}
\vspace{24pt}

\begin{abstract}

  It has recently been suggested that gauge invariance forbids the
  coupling of a massive color-octet vector meson to two gluons.  While
  this is true for operators in an effective Lagrangian of dimension
  four or less, we demonstrate that dimension six interactions will
  lead to such couplings. In the case of technicolor, the result is a
  technirho-gluon-gluon coupling comparable to the naive vector meson
  dominance estimate, but with a substantial uncertainty.  This has
  implications for several recent studies of technicolor
  phenomenology.

\pagestyle{empty}
\end{abstract}
\end{titlepage}

%%%%%%%%%%%%%%%%%%%%%%%%%%%%%%%%
\setcounter{section}{0}

%\section{Introduction}
\setcounter{equation}{0}

Modern technicolor \cite{Weinberg:1979bn,Susskind:1979ms}
theories\footnote{For recent reviews of theories of dynamical electroweak
  symmetry breaking, see
  \protect\cite{Lane:2000st,Lane:2000pa,Chivukula:2000mb} and references
  therein.} incorporate many technifermions in order to produce a theory in
which the technicolor coupling runs very slowly, or ``walks''. Such a
behavior can produce an enhancement of the technicolor condensate
\cite{Holdom:1981rm,Holdom:1985sk,Yamawaki:1986zg,Appelquist:1986an,Appelquist:1987tr,Appelquist:1987fc}
and ordinary fermion masses, thereby mitigating potentially dangerous
flavor-changing neutral currents from extended technicolor interactions
\cite{Eichten:1979ah,Dimopoulos:1979es}. In such a theory the technicolor
scale and, in particular, the lightest technivector meson resonances (the
analogs of the $\rho$ and $\omega$ in QCD) may be much lighter than 1 TeV
\cite{Lane:1989ej,Lane:1991qh,Eichten:1996xn,Eichten:1996xm,Eichten:1996dx}.

If some of the technifermions carry color, we expect that there will be
color-octet technivector mesons.  An analysis based on vector meson dominance
\cite{Eichten:1984eu,Lane:1991qh,Eichten:1996xn,Eichten:1996xm} suggests that
the dominant production mechanism for a color-octet technirho at the Tevatron
collider would be through gluon fusion. While several decay mechanisms are
possible, a coupling of the technirho to pairs of gluons would give rise to a
substantial branching ratio to two jets.  CDF has searched for color-octet
vector mesons decaying to dijets \cite{Abe:1995jz,Abe:1997hm} and $b\bar{b}$
\cite{Abe:1998uz}, and (at the 95\% CL) excludes $\rho^8_T$ with masses in
the range 260 -- 480 GeV/c$^2$ for dijet decays and in the range 350 -- 440
GeV/c$^2$ for decays to $b\bar{b}$ .  Also based on vector meson dominance,
CDF excludes color-octet technirho mesons decaying to third-generation
leptoquark technipions \cite{Affolder:2000ny, Abe:1998iq} in the mass range approximately
200 and 700 GeV, depending on the technipion mass.

Zerwekh and Rosenfeld \cite{Zerwekh:2001uq} have recently suggested
that gauge invariance forbids the coupling of a massive color-octet
vector meson to two gluons\footnote{Results on the production of
color-octet quarkonia at hadron colliders imply that this coupling
also vanishes to leading order in a non-relativistic ``techniquark''
model of the technirho meson \protect\cite{Cho:1996vh}. The coupling
does not, however, vanish at higher order in the non-relativistic
expansion, nor is there any reason to believe that a technirho meson
would be well described as a non-relativistic bound state.}.  On this
basis, they have studied the Tevatron's prospects of finding a
color-octet technirho produced by quark/anti-quark annihilation and
decaying to dijets \cite{Zerwekh:2001ur} and of observing pair
production of color-octet techni-eta's \cite{Zerwekh:2001uq}.

The analysis presented by Zerwekh and Rosenfeld implicitly uses the
``hidden local symmetry'' formulation \cite{Bando:1985ej} which is
most appropriate for a massive vector meson that is light compared to
the fundamental dynamical scale of the underlying theory. The result
is slightly more general, however, as can be seen by considering the
terms of dimension four or less in an effective Lagrangian describing
the couplings of a color-octet technirho $\rho^\mu$ ($ = \rho^{a\mu}
\lambda^a$, where $\lambda^a$ are the generators of $SU(3)_C$ ) to
gluons $G^\nu$ ($=G^{a\nu} \lambda^a$)
\beqa
{\cal L}_4 & = & {-1\over 2 g^2}\, \tr\, G^2_{\mu\nu} - {1\over 2}\, \tr\,
\rho^2_{\mu\nu} +\epsilon\, \tr\, G^{\mu\nu} \rho_{\mu\nu} \nonumber \\
&& +i \alpha\, \tr\, [\rho_\mu,\rho_\nu] \rho^{\mu\nu} 
+ i \beta\, \tr\, [\rho_\mu,\rho_\nu]  G^{\mu\nu} + 
\gamma\, \tr\, [\rho_\mu,\rho_\nu]^2~.
\label{4dlag}
\eeqa
In this expression, $g$ is the QCD coupling, the QCD covariant
derivative is $D^\mu = \partial^\mu + i G^\mu$,
\beq 
G^{\mu\nu} \equiv {1\over i}[D^\mu,D^\nu]~ \ \ \ \ , \ \ \ 
\rho^{\mu\nu} \equiv D^\mu\rho^\nu - D^\nu \rho^\mu~,
\eeq
and $\alpha$, $\beta$, and $\gamma$ real-valued couplings in the
effective Lagrangian. The kinetic energy terms are
normalized such that, under an $SU(3)$ QCD gauge transformation $U(x)$,
\beq
G^\mu (x) \to U(x)G^\mu (x) U^\dagger(x) 
-i  U(x)\partial U^\dagger(x)~,
\label{gaugei}
\eeq
and
\beq
\rho^\mu (x) \to U(x) \rho^\mu (x) U^\dagger(x)~.
\label{gaugeii}
\eeq
Note that we have chosen a basis in which the $\rho^\mu$ transforms
homogeneously, and hence there is no mass-mixing between the $\rho$
and the gluon.

The term proportional to $\epsilon$ in eqn. (\ref{4dlag}) appears to
give rise to the gluon-$\rho$ mixing \cite{Kroll:1967it} which is
integral to calculations inspired by vector meson dominance
\cite{Eichten:1984eu,Lane:1991qh,Eichten:1996xn,Eichten:1996xm}.
However, as demonstrated by Zerwekh and Rosenfeld 
\cite{Zerwekh:2001uq}, this coupling is illusory. Consider the field
redefinition
\beq
G^\mu \to G^\mu + \epsilon g^2 \rho^\mu~.
\label{redefine}
\eeq
Such a redefinition, which is consistent with the gauge symmetry of eqns.
(\ref{gaugei}) \& (\ref{gaugeii}), eliminates the kinetic energy mixing. In
the new basis, it is clear that there is no direct coupling between the
$\rho$ and two properly defined gluons, although the constants $\alpha$ --
$\gamma$ are modified.  In addition a coupling of the $\rho$ to quarks is
introduced which is, on shell, identical to that predicted by vector meson
dominance \cite{Kroll:1967it}.  Note that the elimination of the two-gluon
coupling has nothing to do with ``universality'' of the $\rho$ couplings, and
is therefore independent of any assumption of hidden local symmetry.

There is no reason, however, to restrict the analysis to terms of
dimension 4.  As the technirho is a composite state in a
strongly-coupled theory, contributions of higher-dimensions operators
may be expected to contribute.  In analogy with the analysis of
anomalous gluon self-interactions
\cite{Simmons:1989zs,Simmons:1990dh}, we find there are only two independent
and potentially relevant operators with dimension 6:
\begin{eqnarray}
{\cal L}_6 = c_1 {\cal O}_1 &+& c_2 {\cal O}_2 \label{sixd} \\
{\cal O}_1 &=& {1 \over 4\pi \Lambda^2} D^\alpha\rho_{\alpha\beta} 
D_\gamma G^{\gamma \beta} \nonumber \\
{\cal O}_2 &=& {i \over 4\pi \Lambda^2}
f^{ABC} \rho^{\alpha \beta A} G_\alpha^{\gamma B} G_{\beta\gamma}^C~. \nonumber
\end{eqnarray}
Here we have normalized the unknown couplings so that the coefficients
$c_{1,2}$ are expected to be ${\cal O}(1)$ by naive dimensional
analysis \cite{Georgi:1993dw,Manohar:1984md,Weinberg:1979kz} and
$\Lambda$ characterizes the scale of the underlying strong technicolor
interactions. As in the case of anomalous gluon couplings, application
of the equations of motion shows that the first interaction gives rise
only to couplings of on shell $\rho$ particles to quarks.  The second
interaction in eqn. (\ref{sixd}) does give rise to nonvanishing on-shell
couplings of a color-octet vector meson to two gluons.  Applying the
analysis of \cite{Simmons:1990dh} shows that the contributions of
operators with $d \geq 8$ to the $\rho q \bar{q}$ and $\rho G G$
vertices are identical in form to the contributions from the $d=6$
operators, but are multiplied by powers of $m_\rho^2 / \Lambda^2$.

To compare the effects of the dimension-six interactions with our
expectations from vector meson dominance, we calculate the partial
decay width of the $\rho$ into a pair of gluons. The operator ${\cal
O}_2$ yields the partial decay width
\beq \Gamma_6 (\rho \to GG) =  {c^2_2 g^4\over 16\pi (4\pi)^2} {m^5_\rho
  \over \Lambda^4}~.  
\eeq
In the vector meson dominance calculation, one assumes
\cite{Sakurai:notes} a $\rho$ - $G$ coupling 
\beq
{\cal L}_{VMD} = {g m^2_\rho \over g_\rho} \rho^{\mu A} G_\mu^A
\label{mrhog}
\eeq
where $g_\rho$ is analog of the $\rho \to \pi \pi$ coupling
constant in QCD. This coupling yields the partial decay
width \cite{Chivukula:1995dt} 
\beq
\Gamma_{VMD} (\rho \to GG) =  {g^4\over 16\pi g^2_\rho} m_\rho~.
\eeq
Comparing these expressions we find
\beq
{\Gamma_6 \over \Gamma_{VMD}} = {c^2_2} 
\left({g_\rho \over 4 \pi}\right)^2 \left({m_\rho \over \Lambda}\right)^4~.
\label{compare}
\eeq
In terms of dimensional analysis, which implies $g_\rho = {\cal O}(4\pi)$,
$|c_2| = {\cal O}(1)$, and $m_\rho = {\cal O}(\Lambda)$, we see that
these two estimates agree in order of magnitude.  Consistent with (\ref{mrhog})
above, we can estimate the technirho coupling from QCD \footnote{One can 
include
  ``large-$N_{TC}$'' scaling \protect\cite{'tHooft:1974jz} estimates as well,
  in which case $g_\rho = {\cal O}(1/\sqrt{N_{TC}})$ and $\Gamma(\rho\to
    GG)/m_\rho = {\cal O}(N_{TC})$. The $N_{TC}$-counting, however, is the
    same for decays mediated by $O_2$ or by vector meson dominance.} as
  previous phenomenological studies
\cite{Eichten:1984eu,Lane:1991qh,Eichten:1996xn,Eichten:1996xm} have done
\beq
{g^2_{\rho} \over 4\pi} \approx 2.97~.
\label{qcdest}
\eeq
In this case we see that the ratio above is approximately $c^2_2/4$,
for $m_\rho \simeq \Lambda$.

Turning to phenomenology, the major point of interest is the
$\rho^8_T$ production cross section at the Tevatron collider
\cite{Abe:1995jz,Abe:1997hm,Affolder:2000ny,Abe:1998iq,Abe:1998uz}.  For a
relatively narrow $\rho^8_T$, the gluon fusion cross section is
proportional to the two-gluon partial width calculated above. If
$|c_2|$ is of order one, we see that the production through the
operator ${\cal O}_2$ is about one quarter of the naive vector
meson dominance estimate. On the other hand, for $|c_2|$ even as large
as 2, the two estimates are comparable.

At this point, we should mention that the strength of the $\rho-G$ mixing
used in ref. \cite{Lane:1991qh} is a factor of $\sqrt{2}$ too large; this
translates into dijet decay widths and $\rho_8$ production cross-sections
which are a factor of two too large.  As a result, the calculated theoretical
cross-sections which CDF used to set limits on color-octet technirhos
decaying to dijets \cite{Abe:1995jz,Abe:1997hm}, b-jets \cite{Abe:1998uz}, or
leptoquarks \cite{Affolder:2000ny,Abe:1998iq} correspond to the predictions of the $d=6$
operator with coefficient $c_2 \approx 3$ rather than to the predictions of
vector meson dominance ($c_2 \approx 2$).

An additional contribution to technirho production comes from
quark/anti-quark annihilation.  As mentioned earlier, the field redefinition
of eqn. (\ref{redefine}) induces couplings of $\rho^8_T$ to quarks 
\begin{equation}
\frac{g^2}{\sqrt{g_\rho^2 - g^2}}\  \bar{\psi} \rho \psi\ \ .
\end{equation}
Using the
QCD-based estimate (\ref{qcdest}) of $g_\rho$, ref. \cite{Zerwekh:2001ur}
shows that the $p\bar{p}$ cross-section at the Tevatron collider for
producing a technirho from quark/anti-quark annihilation is about one quarter
as large as the estimate from gluon fusion with $c_2 = 3$ in the mass range
$200\,{\rm GeV} < m_\rho < 500\, {\rm GeV}$. On the other hand, if we take
$g_\rho \approx 4\pi$, the cross-section from $q\bar{q}$ would be reduced by
another factor of six.  Given that walking technicolor models are expected to
contain light technirho's precisely because their dynamics differ from those
of QCD, we should be mindful of the uncertainty in our estimate of $g_\rho$.

Overall, then, we expect the total production cross section for a color-octet
technirho at the Tevatron to be comparable to the naive vector meson
dominance estimate, but with an uncertainty of approximately an order of
magnitude. This has implications for several recent phenomenological studies.
In searches for technirhos decaying to dijets or $b\bar{b}$, a reduced
production rate would decrease the Tevatron's reach, but Tevatron run IIb
with 10-20fb$^{-1}$ of data would still be sensitive to the presence of light
technirhos.  In searches for leptoquark technipions decaying primarily to
third-generation fermions, production without resonant enhancement by
color-octet technirhos yields \cite{Abe:1997dn} a lower bound of $M_{LQ} \geq
148$ GeV.  The resonant contribution of color-octet technirhos (corresponding
to $c_2 \approx 3$) was found to increase sensitivity \cite{Affolder:2000ny,Abe:1998iq,Lane:1991qh}; a reduced technirho contribution would presumably yield a
result intermediate between the two.  Finally, \cite{Zerwekh:2001uq} found
the rate of pair-production of color-octet techni-etas in the absence of a
$\rho G G$ coupling to lie below the threshold of visibility; including the
$\rho G G$ coupling might bring light techni-eta's over that threshold.

%%%%%%%%%%%%%%%%%%%%%%%%%%%%%%

\centerline{\bf Acknowledgments}

We thank Ken Lane, Scott Willenbrock, Fabio Maltoni, Ira Rothstein,
and Rob Harris for useful conversations. {\em This work was supported
in part by the Department of Energy under grant DE-FG02-91ER40676, by
the National Science Foundation under grant PHY-0074274, and by the
Radcliffe Institute for Advanced Study.}

%%%%%%%%%%%%%%%%%%%%%%%%%%%%%%%%%%%%%%%%%%%%%%%%%%%%%%%%%%%%%%%%%%%%%

\bibliography{technirho}
\bibliographystyle{h-elsevier}
\end{document}